\documentclass[aip,rsi,reprint,graphicx]{revtex4-1} 
\usepackage{graphicx}
\usepackage{float}

\draft

\begin{document}

\title{Tunable VUV laser based spectrometer for Angle Resolved Photoemission Spectroscopy (ARPES)}



\author{Rui Jiang}
\affiliation{Division of Materials Science and Engineering, Ames Laboratory, Ames, Iowa 50011, USA}
\affiliation{Department of Physics and Astronomy, Iowa State University, Ames, Iowa 50011, USA}

\author{Daixiang Mou}
\affiliation{Division of Materials Science and Engineering, Ames Laboratory, Ames, Iowa 50011, USA}
\affiliation{Department of Physics and Astronomy, Iowa State University, Ames, Iowa 50011, USA}

\author{Yun Wu}
\affiliation{Division of Materials Science and Engineering, Ames Laboratory, Ames, Iowa 50011, USA}
\affiliation{Department of Physics and Astronomy, Iowa State University, Ames, Iowa 50011, USA}

\author{Lunan Huang}
\affiliation{Division of Materials Science and Engineering, Ames Laboratory, Ames, Iowa 50011, USA}
\affiliation{Department of Physics and Astronomy, Iowa State University, Ames, Iowa 50011, USA}

\author{Colin D. McMillen}
\affiliation{Department of Chemistry, Clemson University, Clemson SC 29634}

\author{Joseph Kolis}
\affiliation{Department of Chemistry, Clemson University, Clemson SC 29634}

\author{Henry G. Giesber III}
\affiliation{Advanced Photonic Crystals LLC, Fort Mill, SC 29708}

\author{John J. Egan}
\affiliation{Advanced Photonic Crystals LLC, Fort Mill, SC 29708}

\author{Adam Kaminski}
\affiliation{Division of Materials Science and Engineering, Ames Laboratory, Ames, Iowa 50011, USA}
\affiliation{Department of Physics and Astronomy, Iowa State University, Ames, Iowa 50011, USA}


\date{\today}

\begin{abstract}

We have developed an angle-resolved photoemission spectrometer with tunable VUV laser as a photon source. The photon source is based on the fourth harmonic generation of a near IR beam from a Ti:sapphire laser pumped by a CW green laser and tunable between 5.3eV and 7eV. The most important part of the set-up is a compact, vacuum enclosed fourth harmonic generator based on KBBF crystals, grown hydrothermally in the US. This source can deliver a photon flux of over 10$^{14}$ photon/s. We demonstrate that this energy range is sufficient to measure the k$_z$ dispersion in an iron arsenic high temperature superconductor, which was previously only possible at synchrotron facilities. 

\end{abstract}


\maketitle 

\section{INTRODUCTION}

ARPES is an excellent technique to study the electronic properties of solids \cite{Damascelli,Campuzano}. It plays an important role in condensed matter physics because it can directly measure the Fermi surface, band structure and study details of energy gaps and electronic interactions. Over the last decade and a half, ARPES has matured as an experimental technique due to technological advances in electron analyzers and the arrival of a third generation of synchrotrons. A major driving force behind this effort was the need to understand high profile physical problems such as high temperature superconductivity in the cuprates. 

In recent years, tabletop VUV laser sources have become an attractive alternative to synchrotrons. There are two main classes of laser-based ARPES spectrometers. Tunable high energy, ultrafast lasers that utilize high harmonic generation in the nobel gases were introduced by the Boulder group \cite{Bauer hhg, Rodriguez hhg}. These sources yield time resolved measurements and have a tunable photon energy. However, due to the wide bandwidth required for high harmonic generation, they typically lack good energy resolution and beam intensities are significantly below modern synchrotrons due to limits on amplifier repetition rate. 

The second type of VUV laser photon sources are low energy (6 or 7 eV) and based on the sixth harmonic of Nd doped Vanadate picosecond lasers\cite{Shin laser, Zhou laser} or the fourth harmonic generation of a Ti:Sapphire laser\cite{Dessau laser, Silvestri tof}. This type of spectrometer is based on a KBBF crystal\cite{KBBF1,KBBF2,KBBF3} and was first introduced by the group of Prof. Shin of ISSP, University of Tokyo. Due to the very narrow width of the emission line, these spectrometers have excellent energy resolution and photon flux that far exceeds the beams available at modern synchrotrons. The main drawback to using this type of spectrometer is that the photon energy is fixed. This causes difficulties when choosing a photon energy with favorable matrix elements and limits access in momentum space to a single plane. 

\begin{figure*}
\includegraphics[width=\linewidth]{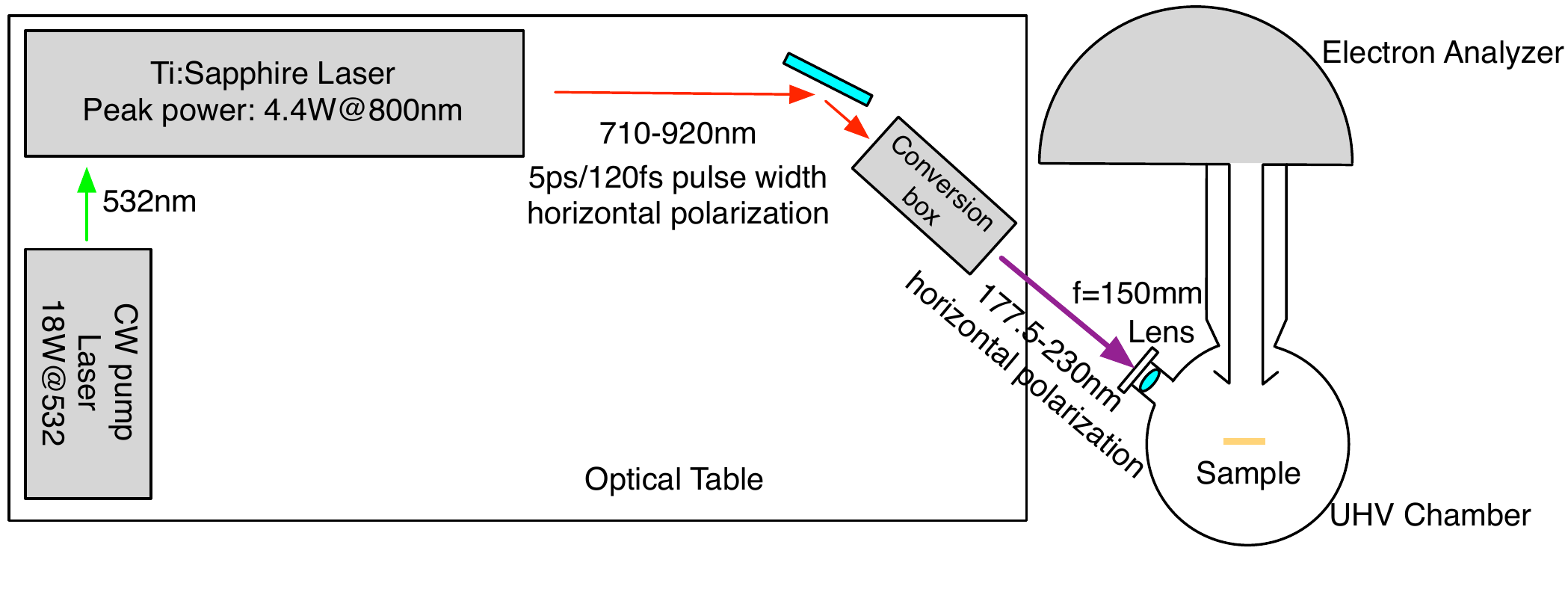}
\caption{Layout of the tunable laser ARPES system. The pump laser, Ti:Sapphire oscillator and FHG conversion box are mounted on a 12" non-magnetic optical table. The electron analyzer and measurement chamber are mounted on an aluminum extrusion frame connected to the optical table.}
\label{SYSTEM}
\end{figure*}

A natural solution to the above limitations is to utilize a tunable laser, where the photon energy can be changed within a certain range. Indeed the design of a tunable VUV laser was described in the literature\cite{FHG} recently, but never used for ARPES studies. Here we describe a complete ARPES system  that utilizes a tunable VUV laser source. The photon source is based on a mode locked Ti:Sapphire commercial laser system Mira HP-D by Coherent Inc. The fundamental IR beam, tunable between 710nm and 1000nm, is first doubled using a standard BBO crystal. We then separate the second harmonic from the remaining fundamental beam and double it again using a KBBF crystal. Due to an embargo on the export of KBBF crystals\cite{Nature china}, we developed a US-based source of these crystals that are grown using a hydrothermal method\cite{kbbf us}. This approach permits the substitution of several alkali metals (such as Rb or Cs) to further tune the mechanical and optical  properties of these crystals. The VUV beam is tunable between 235 nm and 177nm (5.3 eV and 7 eV respectively). It is then focused using a CaF$_2$ lens onto a sample mounted on a cold finger. The Ti:Sapphire laser supports both ps and fs modes of operation. This provides the ability to select either a high energy resolution picosecond mode or femtosecond mode that is favored for time-resolved studies. The typical photon flux in ps mode is $10^{14}$ ph/sec and increases to $10^{15}$ ph/sec in the fs mode. We use a R8000 energy analyzer built by Scienta that was specially tuned for operation at low electron kinetic energies. The lowest usable kinetic energy is around 0.5 eV. The combination of a tunable photon energy range, intrinsically small beam (30 $\mu$m spot size), the ability to perform high energy resolution studies and time resolved measurement makes this an ideal table-top system for use in a small laboratory setting.

We demonstrate that this spectrometer allows access to a significant portion of 3D momentum space  and we use it to measure the 3D dispersion in an iron arsenic high temperature superconductor.

\begin{figure*}
\includegraphics[width=\linewidth]{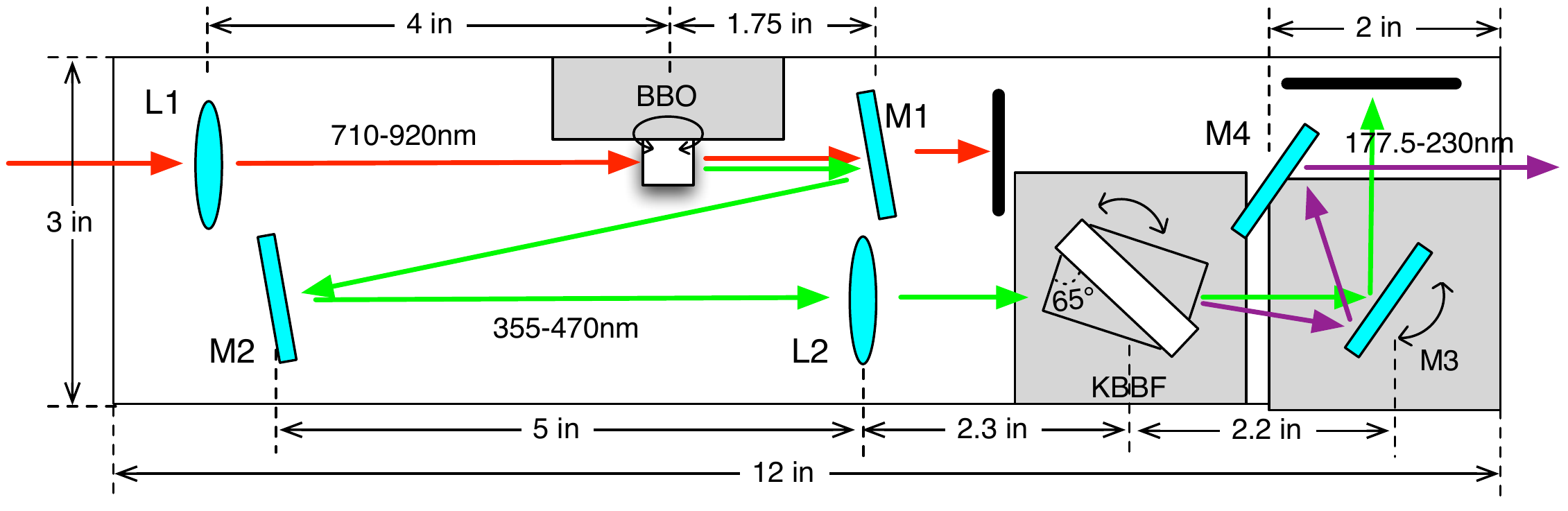}
\caption{Schematic layout of the fourth-harmonic generation. Red, blue and purple arrow indicate the optical path for fundamental light, SHG and FHG. Lens L1 and L2 have a focus length of 200 mm and 50 mm, respectively. Mirror M1 is a cold mirror, only reflecting SHG light.}
 \label{BOX}
 \end{figure*}

\section{INSTRUMENT}

The system consists of a VUV-based photon source, beam delivery optics, experimental chamber and a modern, multiplexing electron analyzer. The schematic layout of the whole system is shown in Fig. \ref{SYSTEM}. Below we describe the key components in detail.

\subsection{Laser}

The fundamental photon beam is produced by a passively mode locked Ti:Sapphire oscillator Mira HP-D supplied by Coherent Inc. The oscillator is pumped by an 18 W CW Verdi laser at 532 nm. The IR beam has power in excess of 3.6W in its tunable range from 710 to 980nm, with maximum power of 5W at 790nm. A high power fundamental beam is necessary in order to produce sufficient photon flux in the VUV range for the high-resolution ARPES measurement. The oscillator also features a dual, ps or fs mode of operation. The picosecond mode has a narrow pulse width (typically $\sim$5ps, corresponding to a $\sim$ 0.4meV bandwidth) that offers better energy resolution, while the femtosecond mode ($\sim$120fs) provides a better time resolution in a pump-probe measurement at the expense of energy resolution. This dual-mode feature allows high energy resolution ARPES measurements and time-resolved ARPES measurements to share the same laser and optical path reducing cost and space requirements. The high, 76MHz repetition rate is essential to avoid problems of space charge that can significantly reduce the energy and momentum resolution of the spectrometer. The setup is very stable and the oscillator can remain mode locked for several days at a time.

\subsection{Conversion box}

Both stages of harmonic generation occur in a conversion box. Because the UV photons below $\sim$200 nm are readily absorbed by oxygen, the conversion box is evacuated down to ($\sim$10 mTorr). The optical components are mounted on a 3$\times$12 inch aluminum plate enclosed in a 4"OD stainless steel tube with 6"CFF flanges on each end. Multiple 2 3/4" viewports, pumping port and electrical feed throughs are attached to the chamber. The schematic layout of the optical components are shown in Fig.\ref{BOX}. The horizontally polarized fundamental beam is first focused by lens L1 (f=200mm) inside a  BBO crystal. Lens L1 is mounted on a translational piezo stage (Newport) that allows compensation for chromatic dispersion. The BBO crystal is mounted on a piezo-driven rotation stage that allows adjustment of the phase matching angle. A dichroic mirror separates the fundamental light from the UV beam after it leaves the BBO crystal. The fundamental light is then dumped on a heat sink or used as a pump. 

The SHG is reflected twice and focused by lens L2 inside a  KBBF crystal assembly. Since the KBBF crystal is very thin ($\sim$1mm), L2 has a very short focal length (f=50 mm) in order to produce higher conversion efficiency. Mirror M2 is mounted on a piezo tilt mount, that allows the adjustment of the beam direction. The phase matching angle of the KBBF crystal for the shortest wavelength is larger than the critical angle, which normally prevents the fourth harmonic from leaving the crystal on the opposite side due to total internal reflection. This problem is solved by sandwiching the KBBF crystal between a pair of CaF2 prisms\cite{2,3,4}. The SHG beam is vertically polarized, therefore the KBBF assembly is mounted on a piezo rotation stage that rotates about a vertical axis. The FHG beam leaves the assembly at an angle of 3$^\circ$ to 17$^\circ$ for wavelengths between 700 nm and 930 nm. In order to maintain the same location of the beam on the sample, we utilize two mirrors, M3 and M4. M3 is mounted on a piezo rotation stage and is used for coarse positioning of the beam onto M4, which in turn is used for fine adjustment of the beam angle. This setup offers a wide tunable range of fourth-harmonic generation from 235nm to 177nm (5.3eV - 7eV), while maintaining the beam at the same position on the sample. The use of very small, vacuum compatible piezo-driven linear, tilt and rotational stages allows for a very compact conversion system that fits inside a 12.5-inch-long 4-inch OD chamber. For easy adjustment, several observation window ports are mounted onto the sides of the chamber. This allows the observation of absolute angles of the stages and optical verification of the beam positions at various stages of conversion.
 
 \begin{figure*}
\includegraphics[width=\linewidth]{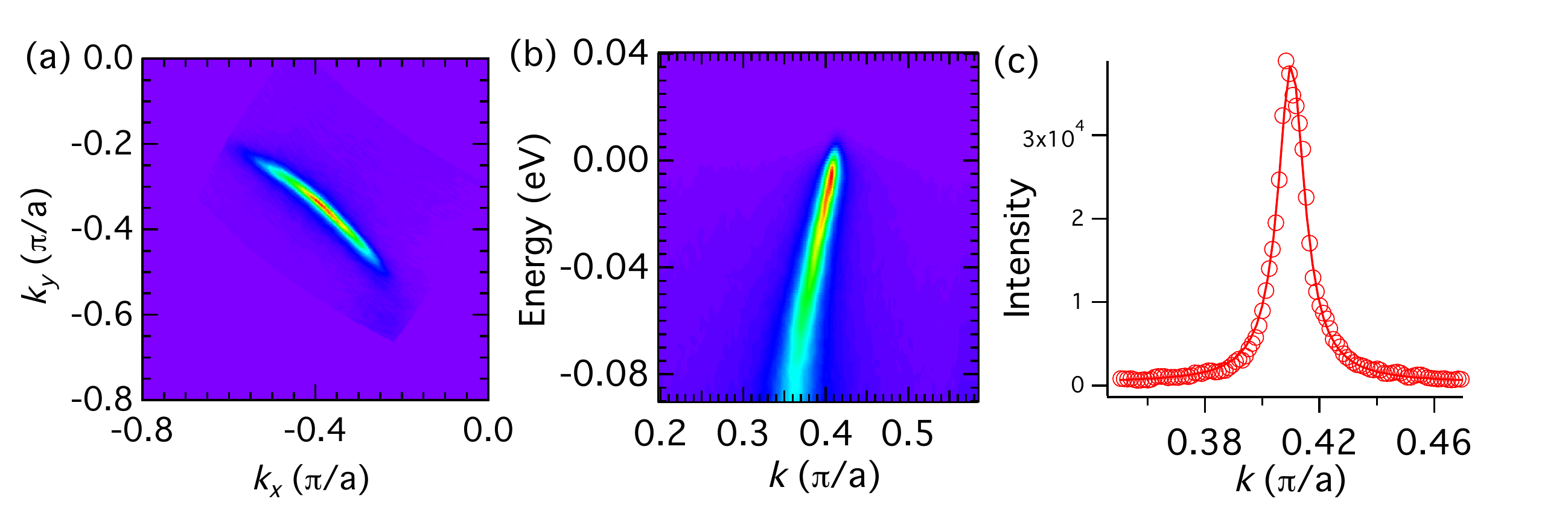}
\caption{Optimal-doped Bi2212 measured with a photon energy of 6.61eV at 13K. (a) Fermi Surface mapping. (b) Intensity along the nodal direction. (c) MDC at the Fermi energy of (b), fitting with Lorentzian function.}
\label{BISCO}
 \end{figure*}

\subsection{Experimental chamber and electron analyzer}

The experimental chamber is a modified standard design supplied by VG Scienta. The walls are manufactured from stainless steel and lined with a mu metal shield to reduce the magnetic field around the sample. The outgoing photoelectrons, especially at low kinetic energies are easily deflected by any magnetic field. This has very adverse effects on the performance of the ARPES spectrometer and will distort the paths of the photoelectrons as they travel from sample to the lens of the analyzer. The mu metal liner mounted inside of the chamber significantly reduces the earth and stray magnetic fields to less than 3 mGauss in the sample-lens area. The chamber is connected to R8000 electron analyzer (supplied by VG Scienta) that was specially tuned for performance at very low kinetic energies. The sample chamber shielding and lens shielding was demagnetized several times to achieve the lowest possible magnetic field inside. When the system is prepared this way it is capable of measuring the angular and energy distribution of photoelectrons down to 0.5 eV without any noticeable distortion of their paths. At the time of writing we use an old design close-cycle refrigerator capable of cooling the sample to only 10K, thus the energy line widths are limited by the sample temperature rather than the electron analyzer or the light source. The VUV beam is focused on the sample by a CaF2 lens with a focal length of 150 mm mounted inside the sample chamber .

\section{INSTRUMENT CHARACTERIZATION}

\subsection{Beam properties}

The fundamental beam is generated in a horizontal polarization, tunable between 710nm and 1000 mn with a typical power of 3.6 to 4.6 W. After the second harmonic generation, the UV beam wavelength is 355 nm to 470 nm and its power is typically 0.8 to 1.2 W. The wavelength for the fourth harmonic VUV beam therefore varies between 177 nm and 235 nm, which  corresponds to 7 eV to 5.4 eV. In the picosecond mode, the peak power of the VUV beam is close to 1 mW at 205 nm and more than 20$\mu$W  at 177.5 nm. The power of the VUV beam is somewhat lower than in a pervious work\cite{2}, due to the thinner KBBF crystal and use of ps rather than fs pulses. However, this intensity at a peak pump power of 18W is more than enough for ARPES measurements and is able to saturate the detector even in the lowest pass energy/slit combination. We adjust the pump beam power down to about 60$\%$ of its maximum to avoid saturation problems. The beam size on the sample is around 30$\mu$m, which is essential for measurements on samples with very small areas of flat surface after cleaving.

\subsection{RESULTS}

To demonstrate the very good instrumental resolution and usefulness of phonon energy tunability, we used the instrument to measure ARPES data on the cuprate, Bi$_2$Sr$_2$CuO$_{6+x}$ (Bi2212) and iron arsenic high temperature superconductor BaFe$_{1.82}$Co$_{0.12}$As$_2$. In Fig. \ref{BISCO} we plot the Fermi surface map of Bi2212 at optimal doping, intensity map and momentum distribution curve at E$_f$ measured along the nodal cut. The width of the MDC is a measure of the scattering rate \cite{Valla} broadened by the experimental resolution. We obtain a width of $\sim$ 0.011\AA$^{-1}$, that is comparable to previous results obtained using a fixed photon energy Laser ARPES \cite{Dessau laser, Zhou laser, Shin laser} and is significantly sharper than typical data obtained at synchrotron facilities.

\begin{figure*}
\includegraphics[width=\linewidth]{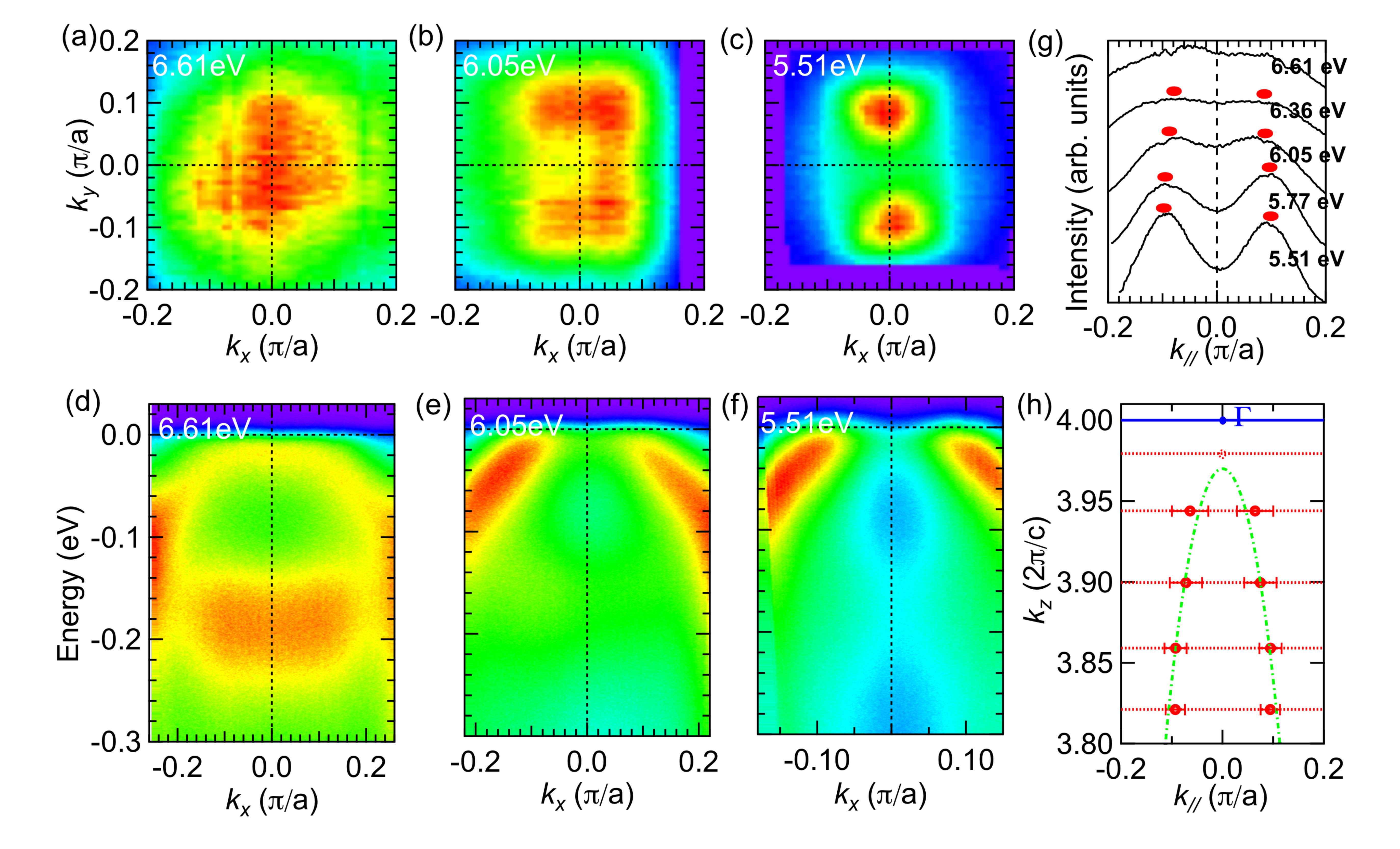}
\caption{Photon energy dependence of BaFe$_{1.82}$Co$_{0.12}$As$_2$ measured at 40K. (a)-(c) Fermi Surface mapping with photon energy 6.61eV, 6.05eV and 5.51eV. (d)-(f) Intensity plot of (a)-(c) along the $\Gamma$-X direction. (g) MDCs at the Fermi level at various photon energies. Red points mark peak positions. (h) $k_z$ dispersion extracted from (g). Solid red circles mark the Fermi crossing with error bar. Dashed red circle shows no Fermi crossing at 6.61eV. Green line is a guide to the eye for $k_z$ dispersion.}
\label{PNICTIDE}
\end{figure*}

There are two key benefits of tuning the photon energy in ARPES measurements. The intensity of the emitted photoelectrons strongly depends on the photon energy due to matrix elements. At a particular photon energy some bands can be therefore too weak to observe. Tuning to a different photon energy can reveal such ``hidden" bands \cite{Bansil,5}. Moreover, the majority of materials have a 3D electronic structure and performing the measurement at a fixed photon energy allows access to only a spherical cut through the 3D momentum space. By tuning the photon energy one can ``change" the radius of this sphere and therefore map a 3D volume in momentum space. As an example, in Fig. \ref{PNICTIDE} we show data that reveals the 3D dispersion of the hole pocket in the iron arsenic high temperature superconductor Co-doped BaFe$_2$As$_2$. The top panels (a-c) show ARPES data measured at E$_f$ for three different photon energies, which is used to reveal the Fermi surface. In the panels below we present binding energy-momentum intensity plots. Areas of high intensity mark the location of the bands. In panel (d) measured at 6.61 eV (which corresponds to a k$_z$ value of 3.98 with an inner potential set to 12eV), the top of the band is located below E$_f$, which signifies the absence of a central hole pocket for this value of momentum. Upon lowering of the photon energy this band moves to higher binding energy and crosses E$_f$ creating a hole pocket (panels e,f). The data is summarized in panel h, where we plot the locations of the Fermi crossings along the z-direction. Indeed the Fermi surface measured at a synchrotron facility at higher photon energies includes an ellipsoidal sheet centered at the Z-point (i. e. the center of the Brillouin zone boundary, $k_z$=3). 

\section{SUMMARY}

We have developed a Laser-based ARPES system with a tunable light source based on the fourth-harmonic generation that has the ability to adjust the photon energy from 5.4eV to 7eV. The system has very good instrumental resolution and a very small beam spot that allows measurements of very small samples. The tunability of the photon energy is provides access to 3D momentum space avoids problems caused by matrix elements. This instrument brings to the small laboratory some of the capabilities previously only available at large synchrotron facilities. In fact the momentum, energy and spatial resolution at low photon energies surpass those available at large synchrotron facilities.

\begin{acknowledgments}
Construction of the instrument was funded by the U.S. Department of Energy (DOE), Office of Science, Basic Energy Sciences. The research was performed at the Ames Laboratory, which is operated for the U.S. DOE by Iowa State University under contract \# DE-AC02-07CH11358. Crystal Growth Development support by Advanced Photonic Crystals, LLC was funded by the National Science Foundation (NSF) SBIR Phase 1 Grant \# IIP-094582 and SBIR Phase II Grant \# IIP-1058055.
\end{acknowledgments}

\end{document}